\documentclass[a4paper,twoside]{article}

\usepackage{subfigure}
\usepackage{calc}
\usepackage{amssymb}
\usepackage{amstext} 
\usepackage{amsthm}
\usepackage[table,xcdraw]{xcolor}
\usepackage[fleqn]{amsmath} 
\usepackage{pslatex}
\usepackage{enumitem}
\usepackage{textcomp}
\usepackage{url}
\usepackage{array,multirow,graphicx}

\usepackage{SCITEPRESS}    

\begin{document}

\title{Requirements Engineering for Global Systems:\\ Cultural, Regulatory and Technical  Aspects
}


 \author{ 
 \authorname{ 
  Maria Spichkova, Heinz Schmidt
  }
  \affiliation{ 
  School of Science, RMIT University\\
  Melbourne, Australia\\
 $\{$maria.spichkova, heinz.schmidt$\}$@rmit.edu.au
  } 
 }

\abstract{
In this paper we present a formal framework for analysis and optimisation of the requirements specifications of systems developed to apply in several countries. 
As different countries typically have different 
regulations/laws as well as different cultural restrictions, the corresponding specific requirements might differ in each particular case. 
Our framework provides a basis for (1) systematic and formal analysis of the diversity and interdependencies within the sets of the requirements, to avoid non-compliance, contradictions and redundancies;
(2) corresponding systematic process for change management in the case of global system development. \footnote{%
Preprint. Accepted to 14th International Conference on Evaluation of Novel Approaches to Software Engineering (ENASE 2019). Final version published by SciTePress.
}
}

\onecolumn 
\maketitle 
\normalsize \vfill 

%===============================================
\section{\uppercase{Introduction}}
\label{sec:introduction}

Software solutions are applied in many areas of our life.  
In many cases,  users of a system have diverse backgrounds, both  cultural and technical. 
The diversity is especially high if the system is developed for application in several organisations or countries. 
In that case, the overall set of requirements expands by the diverse sets of organisation or country specific requirements, regulations and restrictions.  
Moreover, cultural diversity might lead to the diversity of culture-related requirements also within a single organisation or country.  

Requirements engineering (RE) activities have a critical impact on whether the developed system will satisfy user needs as well as regulations and laws of the countries/organisations, where the system will be applied. RE activities  provide a basis for all other activities within the software development life cycle, such as testing, design, architecture, etc., and the errors within them are a major cause of the issues with the delivery of the product on time as well as of the budget overruns, see e.g., \cite{vanLamsweerde:2008,Pretschner:2007,Sawyer2007}. 
Thus, the task is already complicated even when conducting RE activities for a system that is developed for application within a single country or organisation. 
When the system should rely on the standards, legal regulations, cultural aspects, etc. that are not uniform, 
a corresponding solution is required to deal with the related issues in a systematic and scalable way, see e.g., \cite{prikladnicki2003global}. 
%\

A number of studies demonstrated that the cultural diversity has to be taken into account to make the system sustainable and applicable in a global context, see \cite{alsanoosy2018enase,alsanoosy2018cultural,Borchers,govender2016influence,shah2012studying}.
In the proposed approach, we investigate how to manage the diversity of cultural and technical aspects (as well as the correlations between them). 
 
 The core goals of the proposed framework are (1)~to optimise the process of requirements specification and the corresponding change management, as well as 
(2) to ensure that the  system requirements are fulfilled in a global development  context, where also diversity in the cultural and regulatory requirements is taken into account. 
The framework provides methodological structuring of the requirements for the geographically distributed product development and application.
The purposed approach will  
\begin{itemize}
\item 
help to analyse the relations between requirements formally,
\item
facilitate the tracing of requirements' changes  in a global context, and  
\item
provide an input for the TOPSIS (Technique for
Order of Preference by Similarity to Ideal Solution, see \cite{Mairiza2013,topsis2014}), which would allow to identify the most
preferable solutions with respect to the conflicting requirements.
 \end{itemize}

\emph{Outline:}
The rest of the paper is organised as follows. 
Sections \ref{sec:framework} presents the proposed framework. 
Related work is introduced in Section~\ref{sec:related}. 
Section~\ref{sec:conclusions} summarises the paper and the future work directions.

%===============================================

\begin{figure*}[!ht]
  
  \centering
   \includegraphics[scale=0.5]{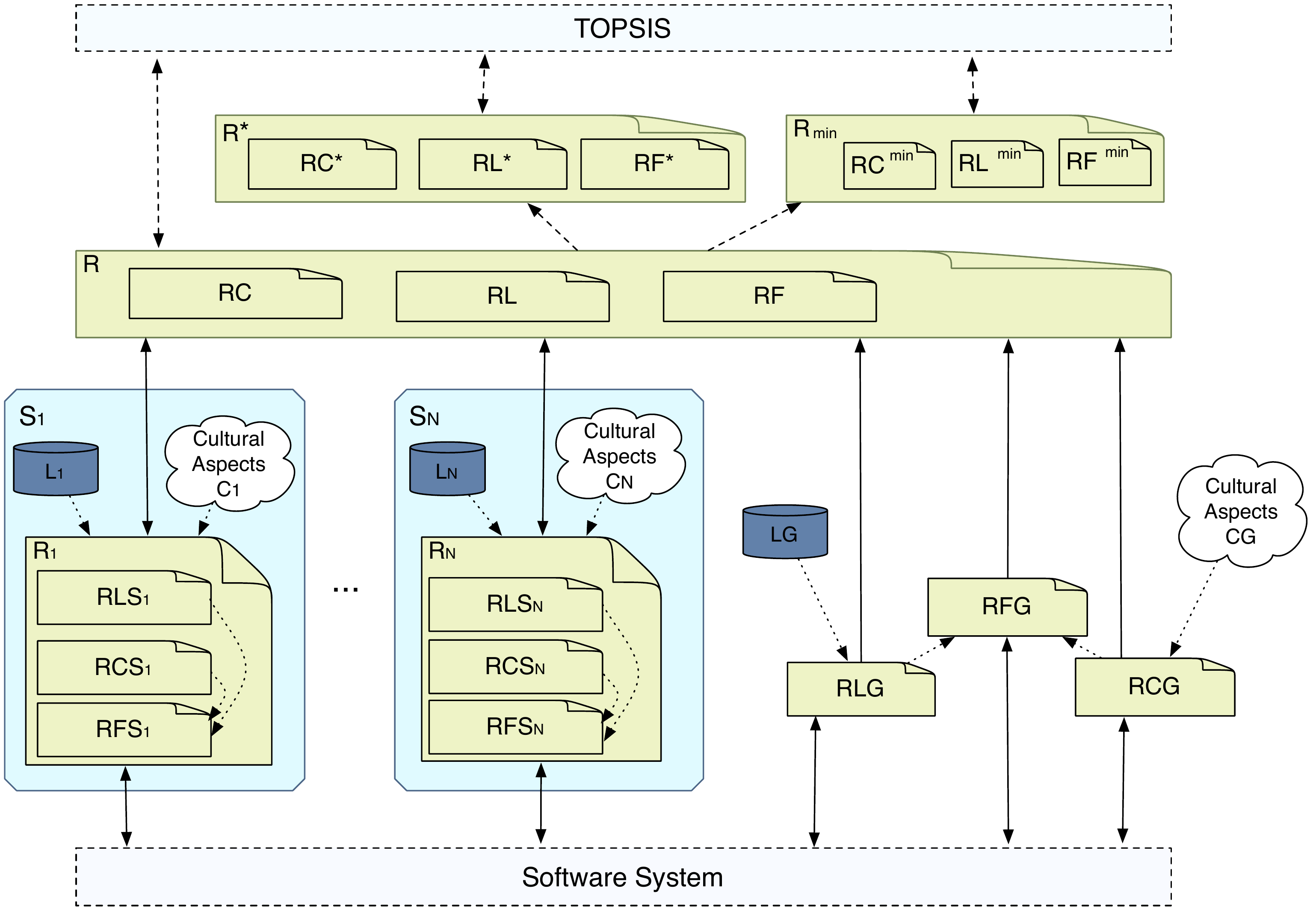} \\ ~
  \caption{% 
  RE framework: Requirements structuring based on the cultural, legal and technical aspects 
  }
  \label{fig:LRset_eL}
 \end{figure*}

%===============================================
\section{\uppercase{Covering The Diversity}}
\label{sec:framework}

In the case of global development, the software system requirements have to
we have to cover not only the technical aspects, but also aspects related to the diversity in culture and country-specific regulations.  
Figure~\ref{fig:LRset_eL} presents a logical architecture of the proposed framework for methodological requirements structuring based on the cultural, legal and technical aspects.

Let us assume that we have to develop a software system  
for application in $N$ countries (or states, organisations, etc.), which we denote $S_{1}, \dots, S_{N}$. 
With each country/state/organisation $S_{i}$, $1 \le i \le N$,  we associate  
\begin{itemize}
\item  
set $L_{i}$ of regulations/laws,   and
\item % 
set $C_{i}$ of cultural influences. 
\item
set $R_{i}$ of functional and nonfunctional requirements to be valid for the country $S_{i}$, which depends on the sets $Reg_{i}$ and $C_{i}$.
\end{itemize}
The complete set of requirements   is then defined by
\[
R  = \bigcup_{j=1}^{N} R_j  
\]
$R$ might contain inconsistencies, i.e. some requirements $R$ might contradict to each other, which should be identified on the early development phases. 
Thus, 
in the case $i \neq j$, we might have a situation where
\begin{itemize}
    \item $L_{i} \neq L_{j}$, and/or
    \item $C_{i} \neq C_{j}$
\end{itemize}
which will also imply
 $R_i  \neq R_j$. 
This also means that we can divide the sets $L_{i}$ and $C_{i}$ into two subsets each to represent 
\begin{itemize}
    \item 
    general components, i.e., regulations/laws and cultural influences common for all countries $S_{i}$, $1 \le i \le N$:
        \begin{itemize}
            \item 
             $LG_{i}$, where  $LG_{i} = LG_{j}$ for any $1 \le i,j \le N$. 
             \item
             $CG_{i}$,  where  $CG_{i} = CG_{j}$ for any $1 \le i,j \le N$. 
        \end{itemize}
        In both cases, we can also omit the bottom index for simplicity and denote the corresponding sets by $LG$ and $CG$.
    \item 
    specific components, i.e., regulations/laws and cultural influences that are specific for some of the countries $S_{i}$, $1 \le i \le N$:        \begin{itemize}
            \item 
             $LS_{i}$, so that for all sets $LS_{i}$, $1 \le i \le N$, holds  
             \[
             \forall x \in LS_{i}.~ \exists 1 \le i \le N.~ \exists y \in LS_{j}.~ contr(x,y)
             \]   
             \item
             $CS_{i}$, so that for all sets $CS_{i}$, $1 \le i \le N$, holds  
             \[
             \forall x \in CS_{i}.~ \exists 1 \le i \le N.~ \exists y \in CS_{j}.~ contr(x,y)
             \] 
             The predicated $contr(x,y)$ denotes the fact that there is a contradiction between $x$ and $y$, which are in two regulations/laws or cultural influences.
        \end{itemize}
\end{itemize}
Thus, for all $1 \le i \le N$ holds
\[
    L_i = LG \cup LS_i
\]
\[
    C_i = CG \cup CS_i
\]
Respectively, the set of functional and nonfunctional requirements $R_i$ to be valid for the country $S_{i}$ can be divided in three (disjoint) subsets. 
\begin{itemize} 
\item 
set $RL_i$ of regulations/laws-based requirements, where some requirements might be  country-specific in the case the corresponding regulations/laws are country-specific. Thus, we can specify it by
\begin{equation}\label{e1}
RL_i =
 RLG\ \cup\ RLS_i 
 \end{equation}
 where
 \begin{itemize}
     \item 
      $RLG$ is a subset of regulations/laws-based requirements elaborated on the basis of $LG$, i.e.,
       \[
        \forall r \in RLG.~ elaboratedFrom(r, RLG)
         \]
      \item
       $RLS_i$ is a subset elaborated on the baseis of $LS_i$, i.e.,
              \[
        \forall r \in RLS_i.~ elaboratedFrom(r, RLS_i)
         \]
 \end{itemize}
\item
 set $RC_i$  denotes the requirements reflecting on culture and economics related aspects, which could be country-specific:  
\begin{equation}\label{e3}
RC_i =
RCG\  \cup\ RCS_i 
\end{equation}
 where
 \begin{itemize}
     \item 
      $RCG$ is a subset of culture-based requirements elaborated on the basis of $CG$, i.e.,
       \[
        \forall r \in RCG.~ elaboratedFrom(r, RCG)
         \]
      \item
       $RCS_i$ is a subset elaborated on the basis of $CS_i$, i.e.,
              \[
        \forall r \in RCS_i.~ elaboratedFrom(r, RCS_i)
         \]
 \end{itemize}
\item
 $RF_i$  denotes the functional and non-functional requirements on the system. 
 These requirements do not depend on the cultural aspects or the regulations and laws directly, but might depend on them indirectly, via the restrictions from the requirements  $RL_i$ and $RC_i$. 
 \begin{equation}\label{eF}
RF_i=
RFG\  \cup\ RFS_i
\end{equation}
 \end{itemize}

We have to build the corresponding ontologies and structure the sets of requirements taking into account the country-specific aspects.  
In the case $RC$ contains a requirement  that is a stronger version of another requirement from $RC$, i.e. is a refinement of it (denoted by $\rightsquigarrow$), the weaker versions should be removed. For example, 
$r_1, r_2 \in RC_i$,  $r_1 \neq r_2$, $r_1 \rightsquigarrow r_2$ implies that 
$r_2$ should be removed as redundant.

While analysing the sets of relevant regulations  $L_{1}, \dots, L_{N}$, 
the following options are possible:
 \begin{enumerate}
 \item

$LG = \emptyset$. This means $RLG = \emptyset$, i.e.,
\[
\forall i.~1 \le i \le N.~  RL_i = RLS_i
\]
In this case, we have to analyse all sets $RLS_i$ especially carefully, as it is possible only if
the sets of regulations are completely different  for all countries 
$S_1, \dots, S_N$. This case is very unlikely. \\
Similarly, the case $CG = \emptyset$, where the cultural aspects and restrictions   are completely different  for $S_{1}, \dots, S_{N}$, is also very unlikely. 
\item 
If the sets of applicable regulations/laws are identical  for all countries $S_{1}, \dots, S_{N}$, i.e.,   $L_{1} = \dots = L_{N}$, we have the situation when 
\[
LG = L_1 = \dots = L_N
\]
and
\[
\forall i.~1 \le i \le N.~  LS_i = \emptyset
\]
which also means 
\[
\forall i.~1 \le i \le N.~ RL_i = RLG
\]
If all requirements in $RLG$ do not change over the time (i.e., are static), the case is the simplest one for the software development: 
the system can be developed on the basis of $RLG$ to use it within  $S_{1}, \dots, S_{N}$. 
The same holds for the  sets of cultural aspects and restrictions: if 
\[
CG = C_{1} = \dots = C_{N}
\]
we can develop a software system on the basis of $RCG$ to use it within $S_{1}, \dots, S_{N}$, as the sets of corresponding country-specific cultural aspects is empty
\[
\forall i.~1 \le i \le N.~  CS_i = \emptyset
\]
 \item
If the sets of regulations/laws are not completely identical  for $S_{1}, \dots, S_{N}$, but have some similarities, i.e., 
\[
 LG \neq \emptyset
 \]
and
\[
\forall i.~1 \le i \le N.~  RC_i = RCS_i\ \cup RCG_i
\]
where 
\[
\exists j.~1 \le j \le N.~  RCG_j  \neq \emptyset 
\]
This is the most common case for the sets of regulations/laws, 
as well as  for the sets of cultural aspects and restrictions, where $CL \neq \emptyset$ and there are differences in the cultural aspects.
\\ 
If these requirements are static (which could be the case for culture-influenced requirements, but  hardly can be assumed for regulations/laws), 
a component-based solution would especially efficient: 
 the components implementing the requirements out of the set $RCG$ 
 can be separated from 
the components implementing   $RCS_1, \dots,  RCS_N$.
As the regulations/laws are typically a subject to change, it is risky to assume that the set $RLG$ will have no changes in the case some $RLS_i$ have changes. 
\end{enumerate}
The following optimisation and reduction of the sets  $RC$  and  $RL$ might increase efficiency of the analysis:
\begin{itemize}
\item
 ${RC_i}^{min}$  and  ${RL_i}^{min}$  denote the sets of cultural and legal requirements, 
 which should be fulfilled by \emph{any} software system (within the corresponding domain) developed for application in the country $S_i$. 
\item
${RC_i}^{*}$ and ${RL_i}^{*}$ denote the strongest sets of the cultural and legal requirements 
for the country $S_i$ (within the corresponding domain). 
We can say that 
these sets are optimisations of 
$RC_i$, and $RL_i$. 
\end{itemize}
~\\
We propose to analyse the sets of requirements based on the optimised views on the sets, i.e., where all redundant (weaker) versions of the requirements are removed, keeping the focus  on the cultural and regulatory/legal aspects. 
As these aspects have different nature, we cannot apply the same strategy to each of them. 
For example, the sets of cultural aspects are usually static, where the regulation/laws are subject to change over time.   %
While identifying $RC^{min}$,  $RL^{min}$ and  $RF^{min}$,   we will analyse which components of the system under development can be reused later. 
This will allow us to have 
\begin{itemize}
    \item an efficient process for the development, 
    \item provide a solution for traceability of the requirements changes that were caused by changes in the regulations in $S_{1}, \dots, S_{N}$. 
\end{itemize}
Thus, if there are some changes in $r \in R_i$, which becomes $r'$ in the new version, the following options are possible:
\begin{enumerate}
\item
The changes affect some $r \in RLS_i$, this  might lead to the following cases:  
    \begin{enumerate}
        \item
        $r'$ is still specific for $S_i$ only, i.e,  only the 
        components implementing the country-specific requirements   $S_{i}$ are affected.
        \item
        $r'$ is now (semantically) identical to the corresponding requirements for all $S_j$, $1 \le j \le N$, $j \neq i$, which means that
        \begin{itemize}
            \item 
            $r'$ should now belong to $RLG$, 
            and all $RLS_1, \dots,  RLS_N$ should be updated respectively;
            \item 
            we might reuse here the corresponding components developed earlier for $S_j$.
        \end{itemize}   
    \end{enumerate}
\item 
The changes affect some $r \in RLG$, this might influence the system as whole. The following cases are possible: 
    \begin{enumerate}
        \item 
        $r'$ is still general for all $S_{1}, \dots, S_{N}$, i.e,  the corresponding
        components implementing the general requirements   are affected. 
        \item
        $r'$ becomes specific for some $S_i$ or for all countries, as not all $RL_i$, $1 \le i \le N$, are affected by these changes. This implies the following 
        \begin{itemize}
            \item 
            we need to revise $RL_1, \dots,  RL_N$ 
            to identify for each of the $S_j$, $1 \le j \le N$, which of the versions -- $r'$  or $r$ --
            should now belong to $RLS_j$;
            \item 
            if $RLS_j$ is now extended by $r$, no changes are required for the components developed for $S_j$;
            \item 
            if $RLS_j$ is now extended by $r'$, the corresponding changes have to be implemented for the components developed for $S_j$.
        \end{itemize}   
    \end{enumerate}
\end{enumerate}
Specification of $RC^{*}$,  $RL^{*}$ and  $RF^{*}$, can provide us a global view on the the system requirements, 
which is not overloaded with the redundant requirements, as all weaker versions are identified and removed.
Respectively, these  sets will provide  an input for the TOPSIS framework to identify the most
preferable solutions with respect to the conflicting requirements. 
On the TOPSIS level, the focus will be on general conflict decision analysis, assuming that the cultural and regulatory diversity issues are already resolved. 

In some cases, we might have even different hierarchy levels to conduct a detailed analysis:
\begin{enumerate}
    \item 
    \emph{Organisational level}, where the organisational regulations and the corresponding cultural aspects have to be take into account;
     \item 
    \emph{State level}, where   
            the state regulations/laws and state-specific cultural aspects have to be taken into account,  
    \item 
    \emph{National level}, where  
        \begin{itemize}
            \item the national regulations/laws and country-specific cultural aspects, and
            \item requirements based on the regulations/laws and cultural aspects of the corresponding states
       \end{itemize}   
       have to be taken into account.
\end{enumerate}
Thus, in each country $S_i$, $1 \ le i \le N$, we might have $M(i)$ states $State_1, \dots, State_{M(i)}$, where $M$ is a mapping from $i$ to the corresponding natural number that specifies the state identifier.

The organisational level might be seen 
\begin{itemize}
    \item[(1)] 
    either as a refinement of a state level, 
    where in each $State_k$, $1 \le k \le M(i)$, we deal with $T(i)$ organisations $Org_1, \dots, Org_{T(i)}$, where $T(i)$ is a mapping from $M(i)$ to the corresponding natural number that specifies the organisation identifier;
    \item[(2)]
    or as a level that is orthogonal to the state and national levels, i.e., we assume that all companies that will be using the product are global.
\end{itemize}
The second option can be used in a very limited number of cases: typically, global companies are presented by their 
country-based units which might differ from each other in the terms of rules, regulations, etc. 
This would imply, that each country-based units can be treated as an organisation.
Thus, the option (1) is more realistic in general.

%===============================================
\section{\uppercase{Related Work}}
\label{sec:related}

Glinz \cite{GlinzRE2007} presented a survey on the existing definitions of non-functional requirements (NFRs). 
The survey also includes a comprehensive discussion of the problems with the
current definitions as well as of promising solutions to overcome these problems. 
In our approach, we analyse NFRs from the side of cultural and legal/ regulatory compliance aspects. 
  
Nekvi et al.  \cite{relaw2011} introduced a compliance meta-model as well as identified a number of key artefacts and
relationships to demonstrate compliance demonstration of the system’s requirements
against  engineering standards and government regulations. 

Several other approaches on compliance validation of requirements we introduced by 
Breaux et al.~\cite{Breaux2008}, Maxwell and Anton~\cite{Relaw2009}, and Siena et al.~\cite{Siena2009}. 

Breaux et al.~\cite{Breaux2015} also elaborated techniques for modelling multi-party data flows requirements   
and verifying the purpose specification as well as limitation principles. 

Yin et al.~\cite{eros2013} proposed an approach for compliance validation of the outcomes of business
processes against outcome-focused regulations.  

Sleimi et al. \cite{RE18Sleimi} proposed a
conceptual model for extraction of  semantic metadata using natural language processing, to provide a basis for the analysis of legal requirements. 
In our future work, we would like to investigate these approaches more deeply, to identify which of them can be incorporated or reused in the proposed framework within the step of analysis $RL_i$ wrt. $L_i$, $RC_i$ wrt. $C_i$, as well as $RF_i$ with respect to $RL_i$ and $RC_i$.

Levy at al.~\cite{levy2018requirements} presented 
a methodology for knowledge management solutions within RE process, which covers both technical and social aspects. 
Spichkova and Schmidt~\cite{reENASE2015} analysed the RE aspects of a geographically distributed architecture in general. 
This analysis was then further refined by Spichkova et al.~\cite{spichkova2015relaw} with the focus on regulatory aspects and variances in compliance. 
In the presented approach we went further, by taking into account cultural aspects as well as providing a formal basis for change management procedure and analysis of interdependencies among requirements, restrictions/laws and cultural aspects. 

Alharthi et al. \cite{alharthi2016individual,alharty2018gender} analysed individual and social (including cultural) requirement aspects of sustainable systems, focusing on educational domain (so-called eLearning systems).

Mairiza et al. \cite{Mairiza2013,topsis2014} introduced the
TOPSIS  framework, which adopts Multi Criteria Decision Analysis approach for NFRs  and 
could assist software developers select the most
preferable design solutions with respect to the conflicting NFR.
TOPSIS does not take into account possible diversity in cultural and regulatory aspects, focusing on general conflict decision analysis. 
In our future work we are going to integrate the TOPSIS in the proposed framework. 

%===========================================================
\section{\uppercase{Conclusions}}
\label{sec:conclusions}

\noindent 
In the case of global system development, we have to take into account that different countries typically have different 
regulations/laws as well as different cultural restrictions, which also implies the corresponding specific requirements might differ in each particular case. 

In this paper, we present a formal framework that allows 
\begin{itemize}
    \item[(1)] to structure and optimise the sets of requirements, as well as 
    \item[(2)] to have a systematic process for change management in the case of global system development, where the diversity in cultural and  legal/ regulatory compliance aspects in  taken into account and analysed especially carefully. 
\end{itemize}
We also discussed in this paper our ongoing work on 
the analysis of interdependencies between the sets of requirements, cultural influences, and regulations/laws.

 \emph{Future Work:}
 In our future work we are going to analyse, which of the discussed in Section~\ref{sec:related} approaches will 
 be the best fit to expand or framework for interdependency and validity analysis of $RL_i$ wrt. $L_i$, $RC_i$ wrt. $C_i$, as well as $RF_i$ with respect to $RL_i$ and $RC_i$.
 
 Another direction of our future work is to integrate the proposed framework with TOPSIS to allow for effective conflict decision analysis.

\bibliographystyle{abbrv}
 
%\bibliography{biblio.bib} 

\end{document}